\documentclass[twocolumn, 11pt]{article}
\usepackage[margin=2cm]{geometry}
\usepackage[T1]{fontenc}
\usepackage{graphicx}
\usepackage{hyperref}
\usepackage{longtable}
\usepackage{array}
\usepackage{tikz}
\usepackage{booktabs}
\usepackage{subcaption}
\usepackage{amsmath}
\usepackage{amssymb}
\usepackage{cleveref}
\usepackage{dsfont}
\usepackage{titlesec}
\titleformat*{\section}{\Large\bfseries}
\usepackage{color}
\hypersetup{colorlinks,
	linkcolor={blue},
	citecolor={blue},
	urlcolor={blue}}

	\title{Collaborative Multi-Radars Tracking by Distributed Auctions}
\def\keywordname{{\bf Keywords:}}
\newcommand{\email}[1]{#1}
\newcommand{\keywords}[1]{\def\and{{\textperiodcentered} }%
	\par\addvspace\baselineskip
	\noindent\keywordname\enspace\ignorespaces#1}
\author{Pierre Larrenie\thanks{THALES SIX GTS France\newline
		4 Av. des Louvresses, 92230, Gennevilliers\newline
		France\newline
		\email{pierre.larrenie@thalesgroup.com}
  }
	\and Cédric L R Buron\thanks{KlaIM, L@bISEN\newline
		33Q avenue du champ de Manœuvre 44470 Carquefou\newline
		France\newline
		\email{cedric.buron@isen-ouest.yncrea.fr}
  } \and
	Frédéric Barbaresco\thanks{THALES LAND \& AIR SYSTEMS\newline
		6 rue de la Verrerie, 92190 Meudon,\newline
		France\newline
    \email{frederic.barbaresco@thalesgroup.com}
}
  }
\date{}

\begin{document}
	\maketitle
\begin{abstract}
  In this paper, we present an algorithm which lies in
  the domain of task allocation for a set of static autonomous radars with
  rotating antennas. It allows a set of radars to allocate in a fully
  decentralized way a set of active tracking tasks according to their
  location, considering that a target can be tracked by several radars, in
  order to improve accuracy with which the target is tracked. The
  allocation algorithm proceeds through a collaborative and fully
  decentralized auction protocol, using a collaborative auction protocol
  (Consensus Based Bundle Auction algorithm). Our algorithm is based on a
  double use of our allocation protocol among the radars. The latter begin
  by allocating targets, then launch a second round of allocation if they
  have resources left, in order to improve accuracy on targets already
  tracked. Our algorithm is also able to adapt to dynamism, \emph{i.e.} to
  take into account the fact that the targets are moving and that the
  radar(s) most suitable for Tracking them changes as the mission
  progresses. To do this, the algorithm is restarted on a regular basis,
  to ensure that a bid made by a radar can decrease when the target moves
  away from it. Since our algorithm is based on collaborative auctions, it
  does not plan the following rounds, assuming that the targets are not
  predictable enough for this. Our algorithm is however based on radars
  capable of anticipating the positions of short-term targets, thanks to a
  Kalman filter. The algorithm will be illustrated based on a multi-radar
  tracking scenario where the radars, autonomous, must follow a set of
  targets in order to reduce the position uncertainty of the targets.
  Standby aspects will not be considered in this scenario. It is assumed
  that the radars can pick up targets in active pursuit, with an area of
  uncertainty corresponding to their distance.
  
\end{abstract}

\section{Presentation of the problem of collaborative multi-radar tracking}\label{presentation-of-the-problem-of-collaborative-multi-radar-tracking}

It is considered that each radar has a 2-dimensional frame in a polar
coordinate system centered on itself. It is estimated that the influence
of elevation is negligible, so it is not useful to use a 3-dimensional
landmark.

\begin{figure*}[!htb]
  \centering
  \includegraphics[width=\textwidth]{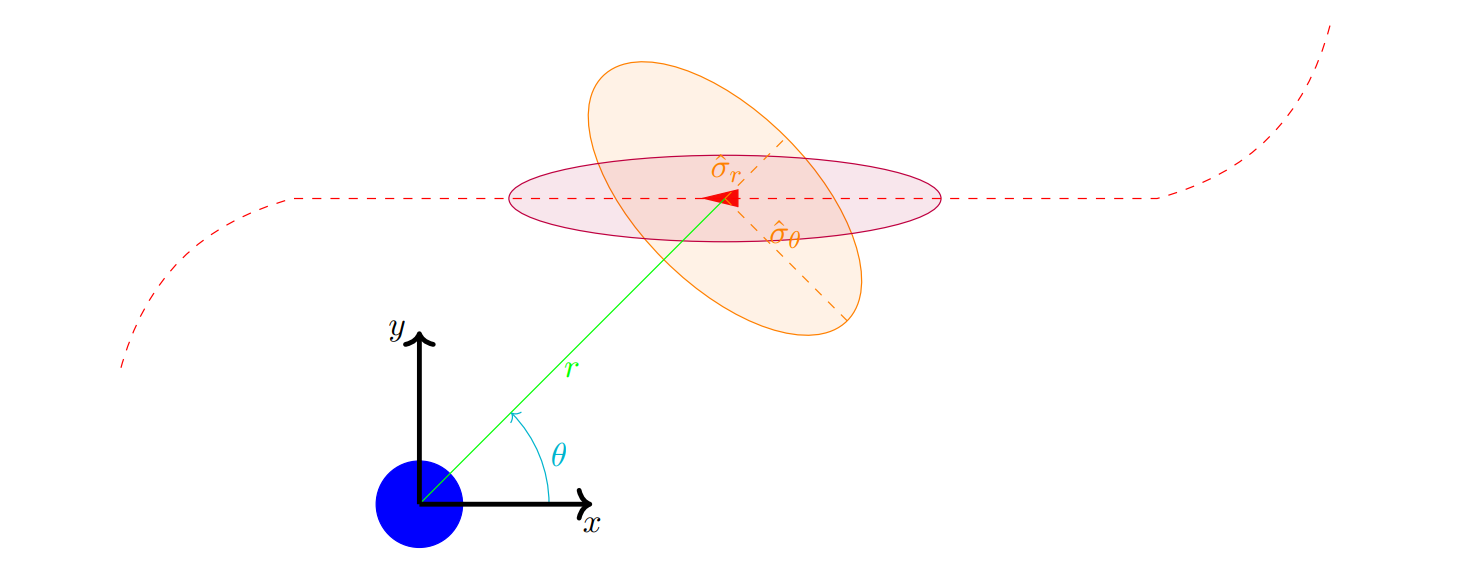}
  \caption{Illustration of uncertainty ellipses during active pursuit}
  \label{fig-uncertainty-ellipses}
\end{figure*}

Each target therefore has a position in the radar reference frame
determined by its distance, denoted \(\) , and its azimuth (polar angle),
denoted \(\theta\) \footnote{A zero angle corresponding to ``North''}.
The precision of the measurement made by the radar is noted:
\(\sigma_{r}\)  in distance and \(\sigma_{\theta}\)  in azimuth. The
resulting measurement uncertainty is represented as an ellipse. The
measure itself corresponds to a\footnote{Corresponding to the Cartesian
  coordinates of the target in a global coordinate system determined by
  a control center}centered 2-dimensional Gaussian random variable of
covariance \(K = R(\theta)\begin{pmatrix}
  \sigma_{r} & 0 \\
  0 & \sigma_{\theta} \\
\end{pmatrix}\) , where represents the \(R\left( \theta \right)\)  angle
\(\theta\) rotation matrix. During active tracking, the aim is therefore
to anticipate the next measurement to be performed on the target given
the past of the target and its current position thanks to the use of a
Kalman filter. As for the measurement, there is also a prediction
uncertainty. This is summarized in \cref{fig-uncertainty-ellipses} with the measurement
uncertainty in orange and the prediction uncertainty associated with the
tracking of the target by the radar in purple.

The signal received by the radar is assumed to be subject to Gaussian
white noise. The signal to noise ratio (S/N)\footnote{The S/N (Signal to
  Noise Ratio) corresponds to the ratio between the noise power of the
  useful signal and the power of ambient noise.} is assumed to be constant. The value was set
at 13, which corresponds to a common value in practice. The S/N will
influence the standard deviation of the measurement. The S/N will in
fact correspond to the quality of the output desired by the user. For a
given S/N it is possible to choose the parameters such as the wavelength
to be transmitted or the transmission power, etc.

We first start by formalizing the problem we are dealing with. We are
interested in the multi-radar target allocation problem. We first
describe the radar model. We then formalize the single-radar allocation
problem, \emph{i.e.} the problem where each target is tracked by at most
one radar, and finally the complete multi-radar allocation problem,
where each target can be tracked by up to two radars.

\hypertarget{first-problem-mono-sensor-allocation}{%
  \section{First problem: mono-sensor
    allocation}\label{first-problem-mono-sensor-allocation}}

It may be noted that the model described below corresponds exactly to a
variant of a backpack problem. This problem can be formalized as
follows:

\[\left( P1 \right):\left\{ \begin{aligned}
  \max\sum_{i,j}{c_{ij} \cdot x_{ij}} \\
  \text{s.t.:} \\
  \sum_{i}x_{ij} \leq 1, & \forall j\in \mathcal{J} & (C1) \\
  \sum_{j}\gamma_{{ij}} \cdot x_{{ij}} \leq L_{t_{i}}, & \forall i\in\mathcal{I} & (L) \\
  x_{{ij}} \in \left\{ 0,1 \right\}, & \forall\left( i,j \right)\in\mathcal{I} \times \mathcal{J}  \\
\end{aligned} \right.\]


\paragraph{Sets :}

\begin{itemize}
  \item
  \(\mathcal{I}\) : Set of cardinality radars\(|\mathcal{I}| = N_{u}\) 
  \item
  \(\mathcal{J}\) : Set of cardinality tasks\(|\mathcal{J}| = N_{t}\) 
\end{itemize}

Note that we are placed here, in the framework
\(\mathcal{I} \ll \mathcal{J}\).

\paragraph{The variables :}

\begin{itemize}
  \item
  \(c_{{ij}}\) : Corresponds to the utility that the radar \(i\) 
  provides to the system if it handles the task \(j\).
  \(c_{{ij}}\) is of the following form, with
  \(V(\mathcal{E}_{{ij}})\)  the surface of the ellipse
  \(\mathcal{E}_{{ij}}\) described by the Kalman filter matrix
  \(P_{{ij}}\) of the radar \(i\)  for the target \(j\): 
  $$c_{{ij}} = f(V(\mathcal{E}_{{ij}}))$$
  Note that the function \(f\) is therefore decreasing according to
  \(V(\mathcal{E}_{{ij}})\).
  \item
  \(x_{{ij}}\) : Boolean variable, \(x_{{ij}}\)  equals \(1\) if
  the radar \(i\)  performs the task \(j\) , 0 otherwise.
  \item
  \(\gamma_{{ij}}\) : Cost of the task \(j\)  for the radar \(i\).
  \item
  \({L_{t}}_{i}\) : Radar budget \(i\).
\end{itemize}

\emph{Constraints :}

\begin{itemize}
  \item
  Constraint (C1) implies that a task can only be performed by a single
  radar.
  \item
  The constraint (L) models the load of the radar which must not exceed
  its budget. The \(\gamma_{{ij}}\) are therefore similar to costs.
\end{itemize}

Let be a total of \(|\mathcal{I}| + |\mathcal{J}|\) constraints.

\hypertarget{second-problem-multi-sensor-allocation}{%
  \section{Second problem: multi-sensor
    allocation}\label{second-problem-multi-sensor-allocation}}

\begin{equation*}
  \left( P2 \right):\left\{ \begin{aligned}
    \max\sum_{i,j,k}{c_{{ikj}} \cdot w_{{ikj}}} \\
    \text{s.t.:} \\
    w_{{ikj}} = {x_{M}}_{{ij}} \land x_{O_{{kj}}},\forall\left( i,k \right) \in \mathcal{I}^{2}, \\\forall j\in \mathcal{J}  (A_{{ikj}}) \\
    \sum_{i}w_{{ij}} \leq 1, \forall j\mathcal{\in J}  (C2) \\
    \sum_{j}\gamma_{{ij}}\cdot ({x_{M}}_{{ij}} + x_{O_{{kj}}} - w_{{iij}}) \leq L_{t_{i}}, \\\forall i\in \mathcal{I} (L) \\
    \left( {x_{M}}_{{ij}},x_{O_{{kj}}} \right) \in \left\{ 0,1 \right\}^{2}, \forall\left( i,j \right)\mathcal{\in I \times J} \\
    w_{{ikj}} \in \left\{ 0,1 \right\},\forall\left( i,k \right) \in \mathcal{I}^{2}, \forall j\mathcal{\in J} \\
  \end{aligned} \right.
\end{equation*}

\emph{The variables :}

\begin{itemize}
  \item
  \(c_{ikj}\) : Corresponds to the utility that the radar \(i\) 
  and the radar \(k\)  provide to the system if the radar \(i\)  handles
  the task \(j\)  as a main radar and \(k\)  as an optional radar.
  \(c_{ikj}\)  is of the following form, with
  \(V(\mathcal{E}_{ij})\) (respectively
  \(V(\mathcal{E}_{kj})\)) the surface of the ellipse
  \(\mathcal{E}_{ij}\) ( resp . \(\mathcal{E}_{kj}\))
  described by the matrix \(P_{ij}\)  (resp. \(P_{kj}\)) of
  the Kalman filter of the radar \(i\)  (resp. \(k\)) for the target
  \(j\)  and \(V(\mathcal{E}_{ij} \cap \mathcal{E}_{kj})\) 
  the intersection volume of these two ellipses, as represented on
  \cref{fig-reconciliation-ellipse}:
\end{itemize}

$$c_{ikj} = f(V(\mathcal{E}_{ij})) + \alpha g(V(\mathcal{E}_{ij} \cap \mathcal{E}_{kj}))$$


$$\left\{ \begin{matrix}
  f(V(\mathcal{E}_{ij})) > > \alpha g(V(\mathcal{E}_{ij} \cap \mathcal{E}_{kj})) \\
  \alpha g\left( V\left( \mathcal{E}_{ij} \cap \mathcal{E}_{kj} \right) \right) > \varepsilon_{\min},\varepsilon_{\min} \in \mathds{R}_{*}^{+},\text{si }\mathcal{E}_{ij} \cap \mathcal{E}_{kj} \neq \varnothing \\
\end{matrix} \right.$$

where:

\begin{itemize}
  \item
  \({x_{M}}_{{ij}}\) : Boolean variable, \({x_{M}}_{{ij}}\) 
  equals \(1\) if the radar \(i\)  performs the task \(j\)  as the main
  radar, 0 otherwise.
  \item
  \({x_{O}}_{{ij}}\) : Boolean variable, \({x_{O}}_{{ij}}\) 
  equals \(1\) if the radar \(i\)  performs the task \(j\)  as an optional
  radar, 0 otherwise.
  \item
  \(w_{{ikj}}\) : Boolean variable, \(w_{{ikj}}\)  equals
  \(1\) if the radar \(i\)  performs the task \(j\)  as main radar and the
  radar \(k\)  performs the task \(j\)  as optional radar, 0 otherwise.
\end{itemize}

\emph{Constraints:}
\begin{itemize}
  \item ($A_{ikj}$):Constraints making it possible to interface between the objective function and the other constraints according to the category of the task for
  the radar. The operator $\wedge$ corresponds to the logical \verb!AND! operator. The constraint summarized here is equivalent to the two constraints below:
  \[\begin{matrix}
    & {x_{M}}_{{ij}} + {x_{O}}_{{ij}} - 2w_{{ikj}} \leq 1 \\
    & {x_{M}}_{{ij}} + {x_{O}}_{{ij}} - 2w_{{ikj}} \geq 0 \\
  \end{matrix}
  \]
  It may be interesting to note that if $w_{iij} = 1$ then, it
  is considered that the task is only performed by a
  single main radar.
  \item
  \((C2)\) : This constraint corresponds to the constraint \((C1)\) of the
  first model. It lists all possible combinations of 2 sensors that
  track a target \(j\). There is at most only one combination of sensors
  that can be chosen.
  \item
  \((L)\) : Just like the constraint \((L)\)  of \(\) , this constraint
  models the load of the radar. If the radar is tracking the target as
  main or optional radar, the term between parentheses equals 1,
  otherwise it equals 0 and the load for the task is therefore not
  considered.
\end{itemize}

Let be a total of
\(|\mathcal{I}|^{2}\cdot |\mathcal{J}| + |\mathcal{J}| + |\mathcal{I}|\) constraints.

Note that the present formulation is difficult to generalize to a
coupling of \(n\) sensors, because it requires adding \(n - 1\) additional
indices which would increase the number of constraints and Boolean
variables far too much. Moreover, this would make the problem
insoluble\footnote{Insoluble in a reasonable time. Indeed, the presence
  of \(n\) Boolean variables requires performing an enumeration, ie
  \(2^{n}\) of possibilities.} for a classical solver. However, it is
estimated that in our case, two sensors are more than enough to have a
significant precision on the target.

\hypertarget{adaptation-of-distributed-auctions-algorithms-for-multi-radar-tracking}{%
  \section{Adaptation of distributed auctions algorithms for multi-radar
    tracking}\label{adaptation-of-distributed-auctions-algorithms-for-multi-radar-tracking}}

The approach that we present in this article is based on the use of two
successive CBBA algorithms: the first to make an allocation as the main
radar, the second as an optional radar. In this section, we proceed in
two steps. We first present the CBBA algorithm and the adaptations we
have made to it so that it can consider the specificities of radars. We
then present the general process, which includes the CBBA algorithm,
allows to consider the interactions between radars and the dynamism of
the mission.

\hypertarget{adaptation-of-cbba-to-radars}{%
  \subsection{Adaptation of CBBA to
    radars}\label{adaptation-of-cbba-to-radars}}

The CBBA (\emph{Consensus Based Bundle Auction}) algorithm \cite{choi2009consensus} is
based on the communication between the different radars. The messages
that the radars send to each other can be represented as a set of
vectors. The set of vectors that a radar sends to another one
corresponds to its current knowledge of the system. Using these
messages, and therefore the knowledge of the other speed cameras, each
speed camera can update its knowledge and transmit it in turn. The
transmitted data includes:

\begin{itemize}
  \item
  \(Y\), the vector that contains the known winning bid utility for each
  target. For a radar \(i\) ,
  \(Y = \left( y_{{ij}} \right)_{j \leq |T|}\) 
  \item
  \(Z\), the vector that contains the identity of the radar that won the
  auction for each target. For a radar \(i\) ,
  \(Z = \left( z_{{ij}} \right)_{j \leq |T|}\). Thus, by making
  the link with \(Y\) , if we have \(Z_{j} = i^{*}\) , then we know that
  it is the radar \(i^{*}\)  which made a winning bid of an amount
  \(Y_{j}\) 
  \item
  \(S\), which corresponds to a ``timestamp'' vector, it makes it
  possible to manage conflicts by making it possible to keep the track
  of the contacts between radars. For a radar \(i\), \(S_{i} = \left( s_{{ik}} \right)_{k \leq |A|}\). Each component
  therefore corresponds to the date of the last message received from
  each radar. It therefore makes it possible to determine which of 2
  radars was the last to be in contact with another one and therefore to
  select the most up-to-date information. This vector is updated when an
  agent receives a message. The values of the agent's neighbors are
  updated with the current time, while the values of other agents are
  updated with the most recent time of the agent's neighbors:
\end{itemize}

\[s_{{ik}} = \left\{ \begin{matrix}
  reception\ time \text{ if } k \in i. \\
  \max_{l \in i.}s_{{lk}} \\
\end{matrix} \right.\]

In our case, \emph{for the allocation as main radar}, another vector
will also be sent. This is the vector
\(E = \left( e_{{ij}} \right)_{j \leq |T|}\) that groups the
ellipses leading to the winning bids for each target. This notably makes
it possible to calculate intersections with the latter.

The algorithm is divided into 2 main phases:

\begin{itemize}
  \item
  A selection phase: the radar calculates the utility for each of the
  targets it can process\footnote{\emph{i.e.}, are located within its
    range of perception and move sufficiently quickly along the radial
    axis of the radar}. It then selects targets for which it has utility
  greater than the winning bid, until it cannot add any more due to
  overload, until there are no more targets it can deal, or that the
  proposed bids are too low.
  
  \item
  An update phase: the radar will open the messages it has received
  containing the knowledge of other radars and update its own according to
  rules defined in the form of a table (taken from the article on the
  CBBA), represented on \cref{cbba-update}.
\end{itemize}

The actions shown in the table are as follows:

\begin{itemize}
  \item
  leave unchanged: \(Y_{i}\) and \(Z_{i}\) both remain unchanged
  \item
  update: 
  \(y_{{ij}} \leftarrow y_{{kj}}\) ,
  \(z_{{ij}} \leftarrow z_{{kj}}\) ,\(e_{{ij}} \leftarrow e_{{kj}}\) ,
  \item
  reset: \(y_{{ij}} \leftarrow 0\) , \(z_{{ij}} \leftarrow \emptyset\) ,
  \(e_{{ij}} \leftarrow \emptyset\) 
\end{itemize}

In the static case, when all the radars have the same beliefs, we say
that the consensus is established. That is to say that a distributed
allocation conflict-free could be found; this therefore constitutes the
end of the algorithm.

The CBBA algorithm has a 50\% performance guarantee, \emph{i.e.,} in the
worst case, the global solution obtained is greater than half of the
optimal solution. Certain assumptions are necessary to ensure this
convergence, the main one being the Diminishing Marginal Gain (DMG)
constraint. We note \(b_{i}\)  all the tasks allocated to the radar
\(i\). The DMG constraint is as follows:

\(c_{{ij}} = c_{{ij}}\left( b_{i} \right) \geq c_{{ij}}\left( b_{i} \oplus b \right)\forall b\) 
(DMG)

The constraint (DMG) reflects the fact that the utility for the same
target must be decreasing in the number of tasks.

In our case, we also want to balance the loads of the radars,
\emph{i.e.}, to make sure that a single radar will not take the maximum
of tasks while leaving the others unoccupied. To respect this
constraint, we added a bias:

\[c_{{ij}}^{\text{CBBA}} = \frac{c_{{ij}}}{|b_{i}|}\]

\hypertarget{general-loop-interaction-and-dynamism}{%
  \subsection{General loop, interaction and
    dynamism}\label{general-loop-interaction-and-dynamism}}

The algorithm operates in a closed loop, where the algorithm is executed
at each time step; in particular, the agent makes an allocation as the
main radar, then it makes the allocation as an optional radar if it has
remaining budget. Each allocation is made through a CBBA algorithm, and
therefore includes the two phases of the algorithm (auction and
consensus) explained above. It therefore receives and sends information
on its allocation as main and optional radar at each time step. The
dynamism aspect, which completes the global loop of the algorithm, is
explained below.

The interactive aspect can be understood in the following way: on the
one hand, a radar does not take into consideration the targets which it
follows as main radar in the list of targets which it can take as
secondary radar (that wouldn't matter). On the other hand, the ellipses
sent by the radars are taken into account by the other radars (whether
they are ellipses sent directly by the neighbors of the radar or
transmitted with the winning bids step by step). It is these ellipses
which make it possible to perform the utility calculation for the
allocation as a secondary radar, and which therefore make it possible to
calculate the utility function of the radar for the target. Another
aspect of coupling is how the budget is managed. Indeed, when the agent
arrives at the stage of allocation as main radar, it considers its
budget as all of its remaining budget plus the budget allocated as
optional radar. If ever a new allocation as main radar is possible, it
deallocates the tasks as optional radar with the lowest utility, and
performs a reset as described in the previous section.

To take account of the dynamic aspect of our problem, we only considered
that the consensus was never reached. Thus, all auctions concerning a
target are not considered closed unless the target in question has not
been seen throughout the system for a certain period of time. This delay
must take into account the fact that all the radars of the system are
not in direct communication. After this delay, the radar deletes the
knowledge (which has become useless) that it had on the target and
transmits the information. In the case where the target can be perceived
again by the system, the latter behaves in the same way as for the
appearance of a new target, namely that it creates the knowledge
associated with it. After each allocation step (as main radar and
optional radar), the latter performs the tasks to which it has allocated
itself.

So the general loop (repeated forever) is as follows.

\begin{enumerate}
  \def\labelenumi{\arabic{enumi}.}
  \item
  The radar makes the selection step as the main radar. To do this, it
  calculates the uncertainty ellipses for each target, and its utility
  function; it also applies the Kalman filters of all the targets it is
  already tracking.
  \item
  If it has remaining radar time budget, the radar proceeds to the
  selection step as an optional radar on all the targets that are not
  already tracked as the main radar (with its remaining budget).
  \item
  The radar proceeds to the consensus stage as the primary radar. The
  vectors \(Y,Z,E\) and \(S\) as main radars are updated; vectors are
  sent to neighbors.
  \item
  The radar then proceeds to the consensus stage as an optional radar.
  Vectors \(Y,Z\) and \(S\)  as optional radars are updated; vectors are
  sent to neighbors.
  \item
  The radar tracks the targets it has selected, possibly by applying its
  Kalman filter.
\end{enumerate}

Note that since the radar initiates the tracks at each execution of the
two phases of CBBA, which means that the algorithm does not have time to
converge. In practice, this can result in conflicts. This is the case
for a target on the edge between two radars, and which is increasingly
threatening. In this case, each radar making its decision on a previous
value of the utility of the others, and its own current value, it will
consider that its bid wins the bet. Similarly, a less and less
threatening radar (\emph{i.e.,} whose usefulness decreases) will
potentially be followed by no one, each radar considering that the other
has a better bid than itself. Differentiation from conventional methods

The method we propose calls into question several classic points in
target allocation. If there is indeed a method of target allocation by
auction, our approach proposes a different method, where the role of the
auctioneer is in fact distributed on the radars. The latter therefore
hold two roles at once: that of bidder first, and then that of
auctioneer. This feature is represented by the two phases of the
algorithm: the auction phase and the consensus phase.

Another particularity of our approach is that it allows to allocate a
target to several radars, thus making it possible to take advantage of
the intersection of the ellipses of uncertainties generated by the
different radars, which is not possible in the methods offered in the
state of the art. Finally, we propose a classic auction, and not a
combinatorial auction. While traditional auctions are generally less
efficient than the combinatorial auctions described in the state of the
art, they are also much faster, which allows our approach to achieve a
relatively rapid allocation, particularly in a context where targets are
move quickly.

\begin{figure*}
  \centering
  \begin{subfigure}{.4\textwidth}
    \includegraphics[width=\textwidth]{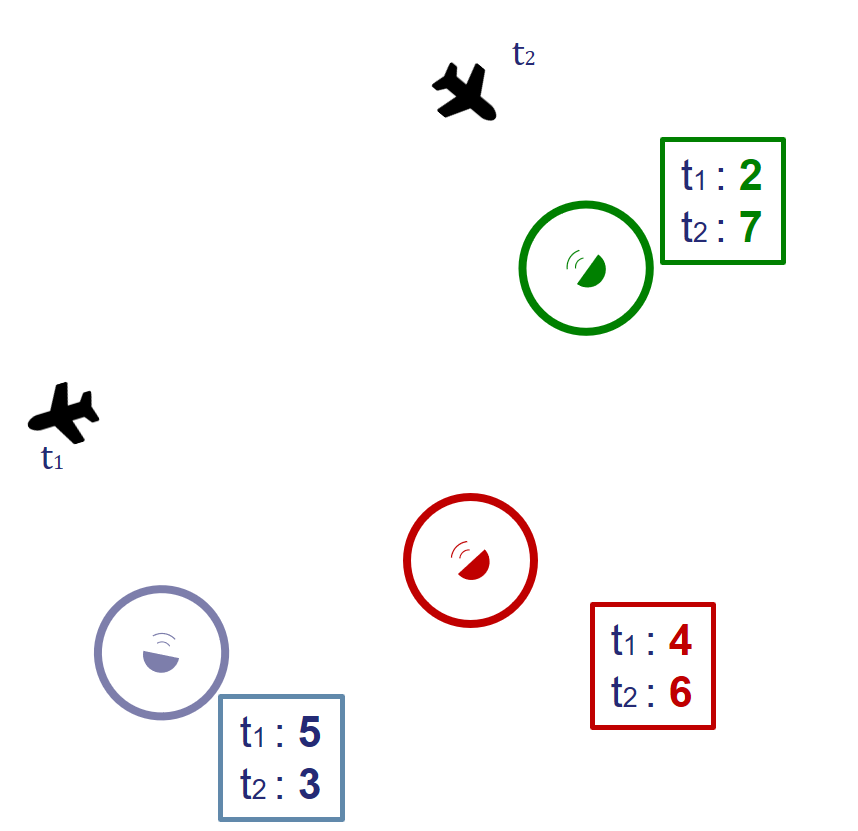}
    \caption{initial allocation}
  \end{subfigure}
  \begin{subfigure}{.4\textwidth}
    \includegraphics[width=\textwidth]{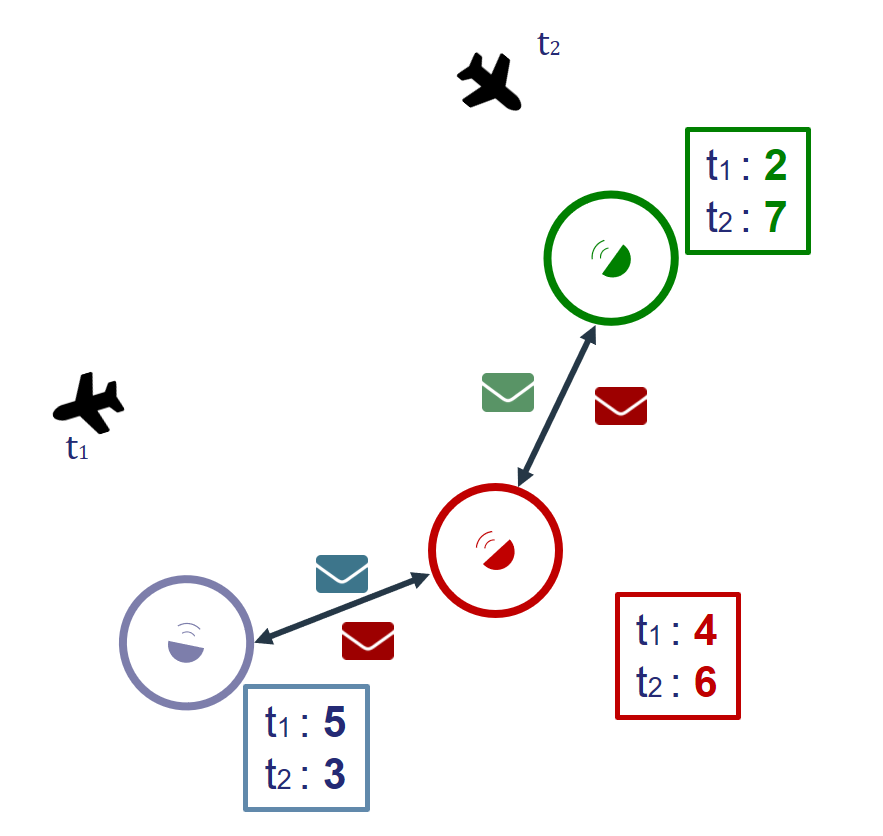}
    \caption{sending messages}
  \end{subfigure}
  \begin{subfigure}{.4\textwidth}
    \includegraphics[width=\textwidth]{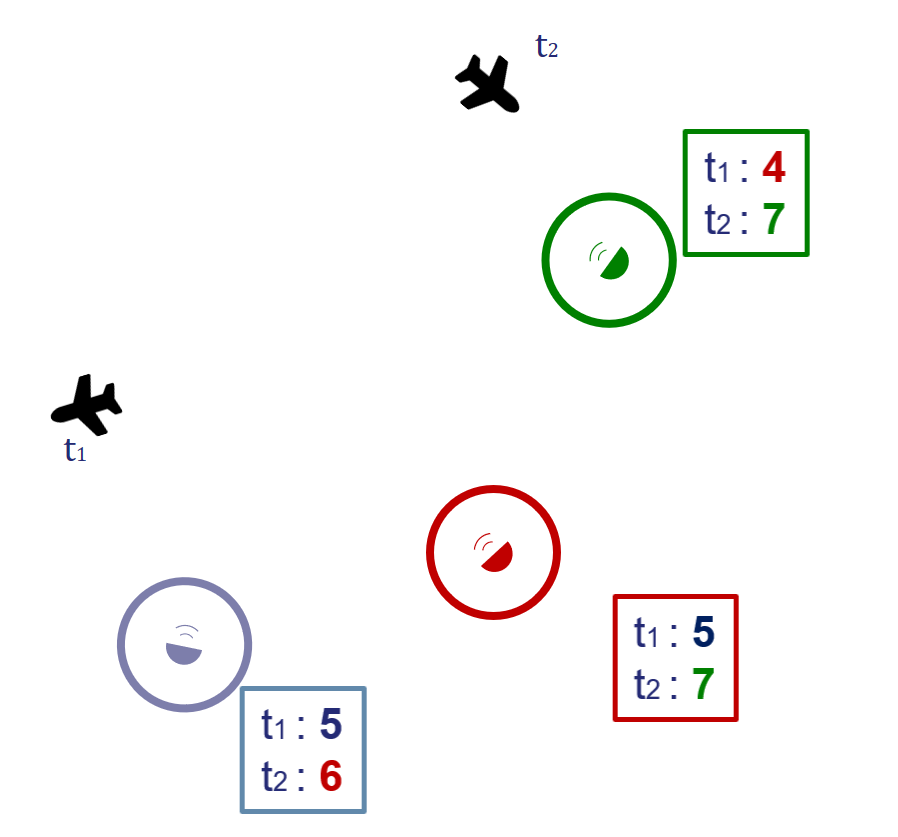}
    \caption{Update (consensus)}
  \end{subfigure}
  \begin{subfigure}{.4\textwidth}
    \includegraphics[width=\textwidth]{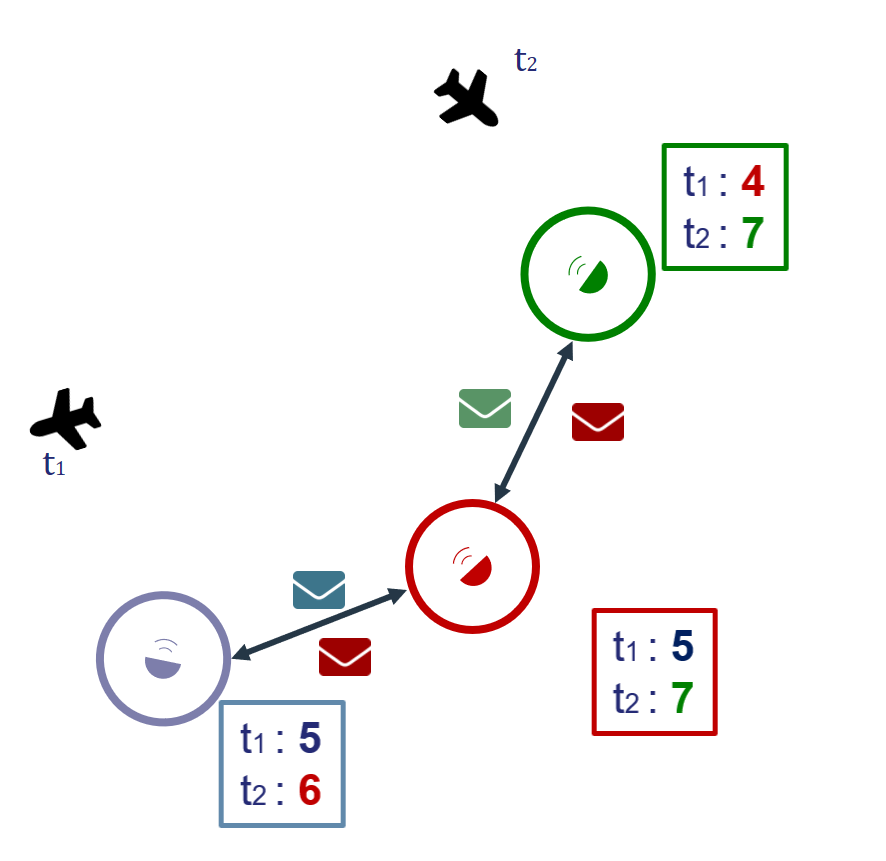}
    \caption{Bidding \& sending messages}
  \end{subfigure}
  \begin{subfigure}{.4\textwidth}
    \includegraphics[width=\textwidth]{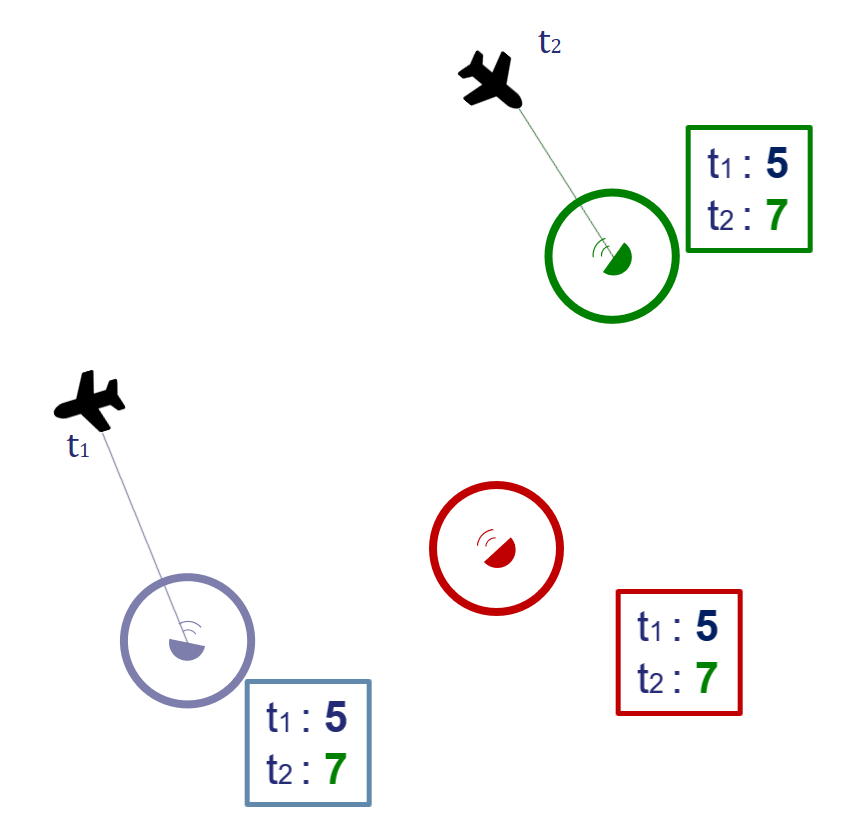}
    \caption{Second consensus stage, consensus reached}
  \end{subfigure}
  \caption{Master Allocation Method Workflow for Two Targets and Three Radars}
\end{figure*}

\hypertarget{algorithm-illustration}{%
  \section{Algorithm illustration}\label{algorithm-illustration}}

Our approach makes it possible to solve the task allocation problem in
an advantageous way: in the case where the radars remain in contact,
even if the communication graph evolves during the mission, this remains
possible provided that the graph of connection of the radars remains
connected by arc. The total decentralization of the algorithm makes it
possible to relax the constraint on the communication graph, which does
not require having a radar in direct communication with all the radars,
the information propagating in the communication graph. In order to make
the allocation in a reasonable time, the calculations of allocation on
the targets are done in an additive manner at the level of each radar,
which makes it possible to reduce the combination thereof. Finally, by
carrying out two sequential auction runs, our algorithm also takes into
account the possible overlapping of uncertainty ellipses, and thus makes
it possible to generate an allocation favoring more precise tracking of
targets when possible, while prioritizing the tracking of as many
targets as possible (depending on radar capability).

The main means of realization is the deployment of the CBBA algorithm on
radars. To proceed with this deployment, each radar is provided with a
part of the implementation of the utility function, which allows it to
determine the utility associated with a target, as well as a decision
function which allows it:

\begin{enumerate}
  \def\labelenumi{\arabic{enumi}.}
  \item
  To bid on the targets it has detected during its standby phase,
  considering its own load limits,
  \item
  To receive messages sent by other radars, 
  \item
  To calculate the winner for each target according to Table 1,
  \item
  Return the updated table.
\end{enumerate}

Radars must also be able to differentiate between first-round allocation
messages -- one that tracks as many targets as possible -- and one that
improves accuracy by generating an intersection of uncertainty ellipses.
Each radar \(i\)  proceeds by trying to maximize
\(\sum_{j}{x_{{ij}} \cdot}c_{{ij}}\) for the main allocation
(resp.\(\sum_{j,k}{w_{{ikj}} \cdot}c_{{ikj}}\)  for the
optional allocation), that is to say the sum of the utilities
corresponding to the targets that it tracks, while taking into account
the information received from the other radars, in particular the bids
made by the latter.

\begin{figure}[!htb]
  \centering
  \includegraphics[width=.4\textwidth]{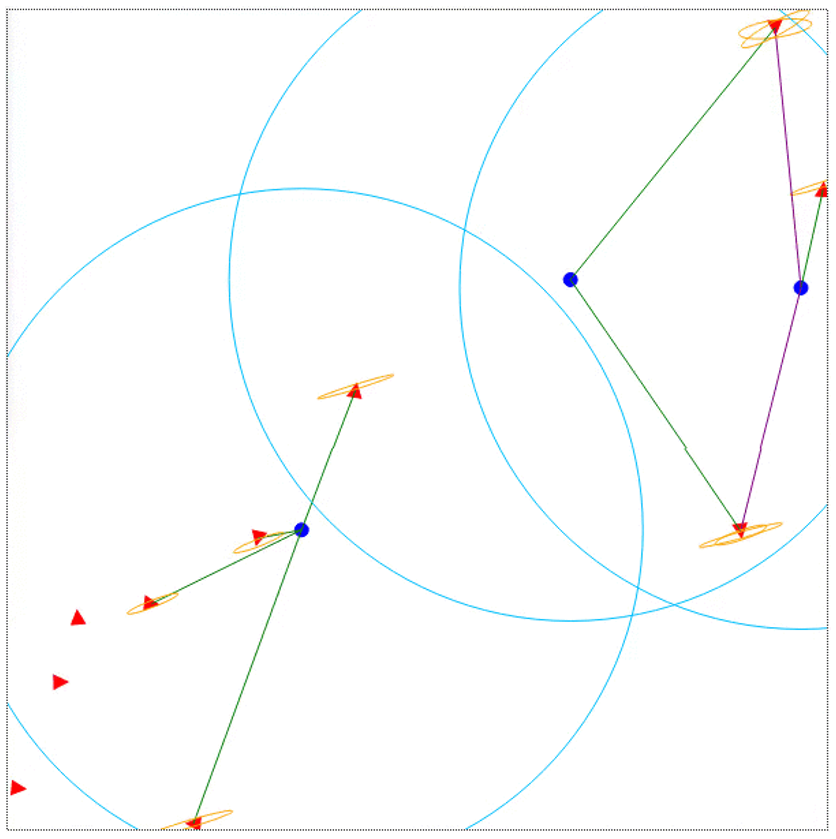}
  \caption{Simulation of the algorithm with 3 radars, the area where they are able to pick up targets in active pursuit, and 10 targets. The main track is in green, the secondary track in purple.}
  \label{simu-radars}
\end{figure}

An example of the way the algorithm works for allocation of the main
radar is represented on \cref{simu-radars}. The implementation must also include
an additional target disambiguation mechanism, making it possible to
identify the targets present at several radars, and in particular a plot
merging algorithm, making it possible to match the targets of the
different radars. This induces the sending of additional information
enabling this operation to be carried out, such as the estimated speed
and position of the targets.

\hypertarget{results}{%
  \section{Results}\label{results}}

The implementation of our model has been performed on the MESA framework
\cite{kazil2020utilizing}, along with the Kalman filter package \cite{laaraiedh2012implementation},
on the same simulator as the one used in \cite{nour2021multi}.
The \cref{simu-radars}
shows the simulator: the radars are represented as blue points while the
targets are represented as red triangles. Main radars following a target
are represented as green lines, optional ones as purple lines. The
uncertainty ellipses are in yellow.

\begin{figure}[!htb]
  \centering
  
  \begin{subfigure}{.5\textwidth}
    \centering
    \includegraphics[width=\textwidth]{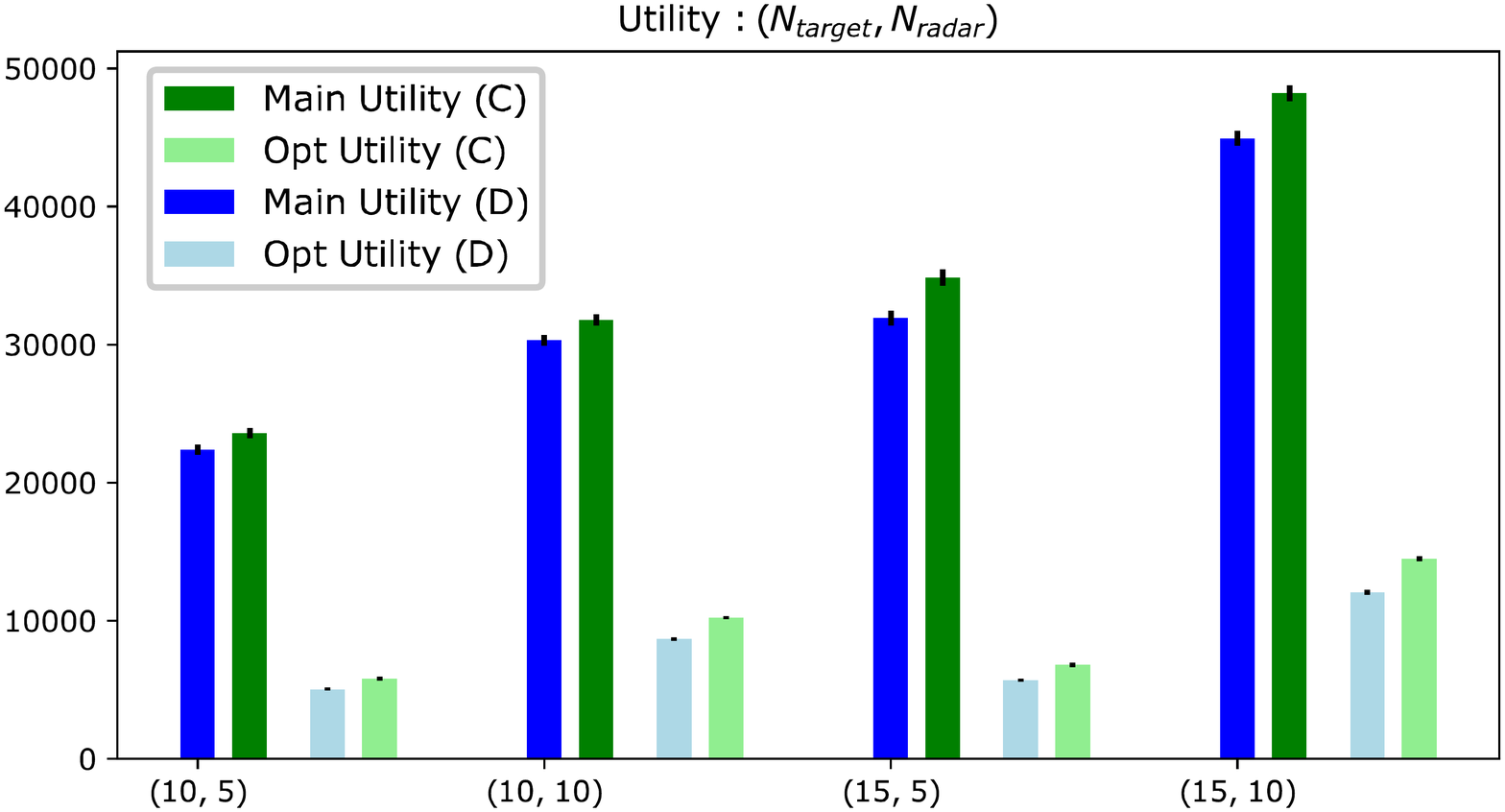}
    \caption{Utility}
  \end{subfigure}
  \begin{subfigure}{.5\textwidth}
    \centering
    \includegraphics[width=\textwidth]{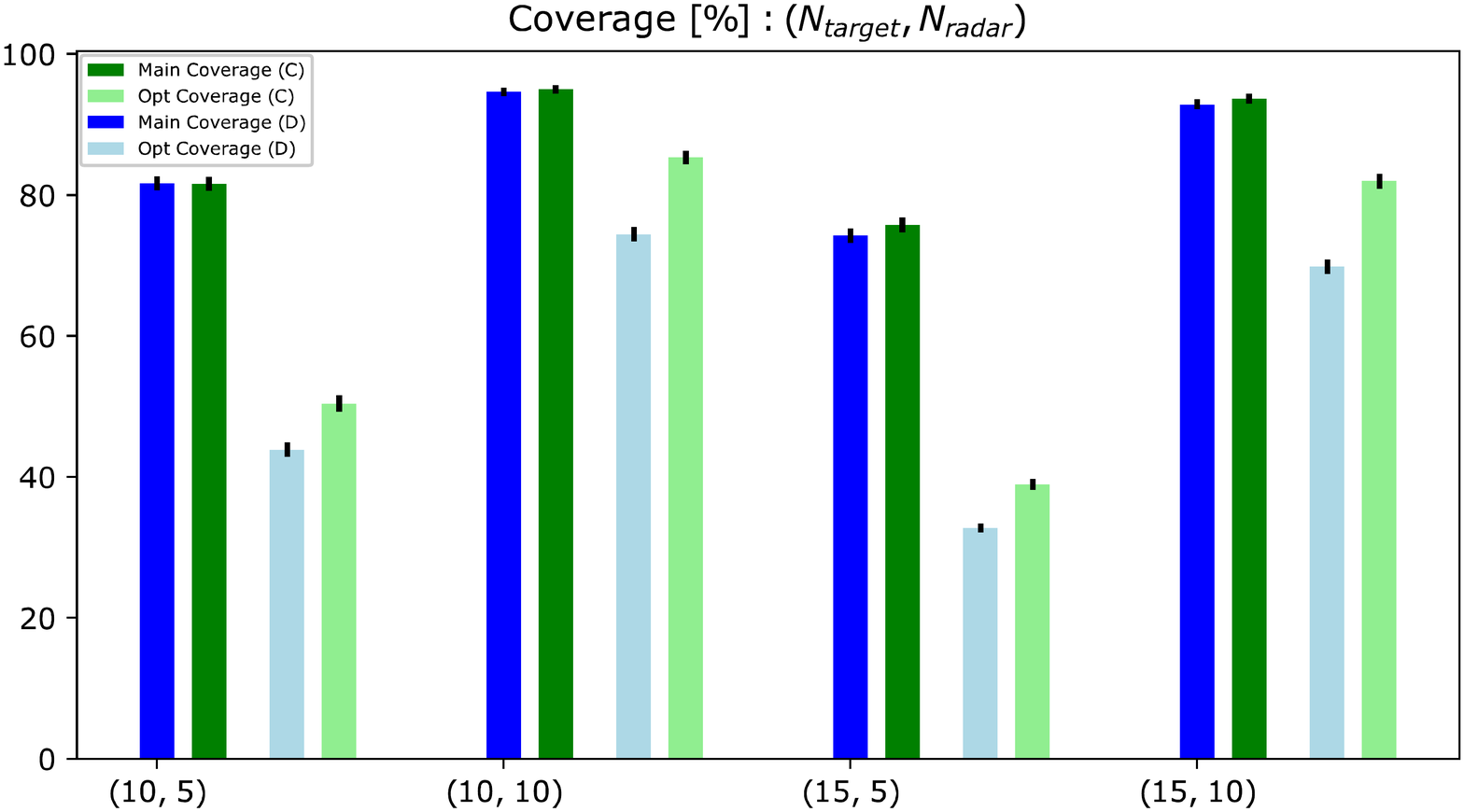}
    \caption{Coverage}
  \end{subfigure}
  \begin{subfigure}{.5\textwidth}
    \centering
    \includegraphics[width=\textwidth]{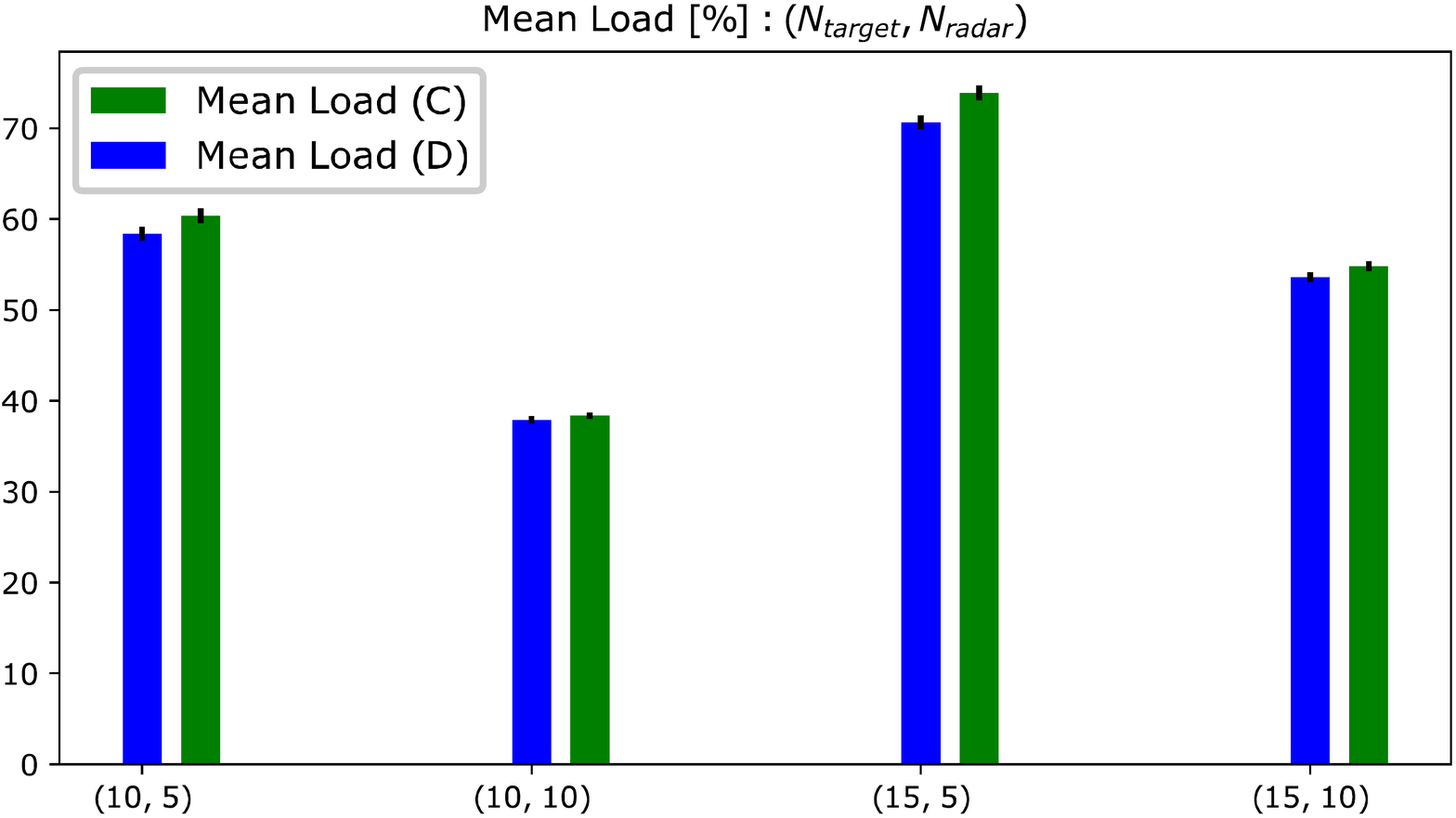}
    \caption{Mean Load of the radars}
  \end{subfigure}
  \caption{Evolution of tracking metrics over scenarios generated randomly}
  \label{fig-results}
\end{figure}

In order to evaluate our work, we compare it to an optimal allocation,
performed with the Coin-OR Branch and Cut tool (CBC). Results are
represented on \cref{fig-results}. The blue (or green) curves correspond to the
decentralized (or centralized) approach. The average value of each of
the simulations and the different scenarios for a certain ``composition''
of Targets and Radars are presented. Each of the ``compositions'' of
Targets and Radars is represented on the x-axis as a tuple (Targets,
Radars). This allows us to compare the performance of the centralized
and decentralized approaches with equivalent configuration. The colors
of the bars correspond to the decentralized (D) or centralized (C)
values, with optional tracking represented in light colors when
relevant. The standard error (in black) is available for each of the
bars of the different graphs.

Regarding the utility, as shown in the previous figure, the centralized
approach obtains results superior to the decentralized approach.
However, it can be seen that here too, for the decentralized approach,
we obtain a utility clearly higher than the theoretical 50\% of the CBBA
algorithm compared to the centralized approach.

Regarding the coverage, we notice that for a constant configuration we
obtain equivalent coverage in ``main'' tracking. The coverage being weaker
for the coverage for the ``optional'' tracking.

Regarding the average load, we observe that for the configurations
studied, there is almost no difference between the centralized and
decentralized approaches. While we would have liked the decentralized
approach should have a much lower load than the centralized approach.

As we can see overall, there are only very small differences between the
centralized and decentralized approaches. This can be interpreted as a
strength for the decentralized approach, because the radar configuration
could then be adaptive (one could suppose that the radars are not fixed
but can move) but also resilient, \emph{i.e.,} the global system could
continue to function normally if a connection is cut, which is not the
case with a centralized approach since there is only one connection with
the control center. Since the load is also more evenly distributed, it
can be assumed that the centralized approach can cope with a ``surprise''
attack without a total re-planning, which is not the case for the
centralized approach. It should be noted that no experiment has been
done with a higher number of radars since the constraint of the cost of
purchasing a set of radars can be very limiting.

\hypertarget{related-works}{%
  \section{Related works}\label{related-works}}

Several works have focused on the use of decentralized approaches for
task allocation for sensor, since the seminal work of Lesser et at.
\cite{lesser2003distributed}. Since then, many approaches have been used. Recently, for real
time use cases, auction methods have gained much interest in the
multi-agent community \cite{gini2017multi,krainin2007application} for their capacity to perform good
allocation in an affordable time. Many methods have been using auctions
since then to allocate tasks in real-time, including to robots, sensors
and radars (see for instance \cite{deliang2017distributed}).

One of the most successful recent algorithms is CBBA \cite{choi2009consensus}. This
algorithm allows to perform the allocation in a fully decentralized way,
the agents acting both as auctioneers and bidders. This algorithm has
since been used several times for sensors \cite{jia2017consensus,sameera2012robust}. However, none of
them has been taking into account the specificities of radars,
\emph{i.e.,} their collaboration through the intersection of their
uncertainty ellipses. Similarly, the challenge of high dynamicity has
been barely studied.

\hypertarget{conclusion}{%
  \section{Conclusion}\label{conclusion}}

In this paper, we presented a novel approach for allocating target to a
team of radars in a totally decentralized way. This approach is based on
a fully decentralized auction algorithm, CBBA. We showed that, when
taking into account the intersection of uncertainty ellipses, the
results of this algorithm is comparable to the centralized optimal
allocation.

Future works include the design of a more generic approach, that could
handle an arbitrary number of radars following the same target. We also
would like to make our approach more dynamic, for instance by including
replanning approaches that have been recently proposed to improve CBBA
\cite{buckman2019partial} and evaluate this approach in the highly dynamic setting
imposed by our use-case.	
	\bibliographystyle{unsrt}
	\bibliography{biblio.bib}
\end{document}